\documentclass[       %
aps,                    
prd,                    
showplace,              
nofootinbib,            
showkeys,               %
preprintnumbers,        %
floatfix]               
{revtex4}               

\usepackage{graphicx,amssymb,amsmath,rotating}

\newcommand{\Bae}{Bethe ansatz equation}
\newcommand{\Baes}{Bethe ansatz equations}

 \allowdisplaybreaks

\begin{document}

\title{Quantum Invariants of the Pairing Hamiltonian}

\author{Y. Pehlivan}
\email{yamac@physics.wisc.edu}

\affiliation{Department of Applied Mathematics, Hali{\c{c}}
University}

\date{\today}
\begin{abstract}
Quantum invariants of the orbit dependent pairing problem are
identified in the limit where the orbits become degenerate. These
quantum invariants are simultaneously diagonalized with the help
of the Bethe ansatz method and a symmetry in their spectra
relating the eigenvalues corresponding to different number of
pairs is discussed. These quantum invariants are analogous to the
well known rational Gaudin magnet Hamiltonians which play the same
role in the reduced pairing case (i.e., orbit independent pairing
with non degenerate energy levels). It is pointed out that
although the reduced pairing and the degenerate cases are opposite
of each other, the Bethe ansatz diagonalization of the invariant
operators in both cases are based on the same algebraic structure
described by the rational Gaudin algebra.
\end{abstract}

\medskip
 \preprint{} \maketitle

\section{Introduction}   

Strong pair correlations are observed in fermionic many body
systems which energetically favor large wave function overlaps.
This phenomenon is known as pairing and it plays an important role
in our understanding of many body physics (See Ref.
\cite{Dean:2002zx} for a review). Historically, the physical
significance of pairing was first realized with the microscopic
theory of superconductivity developed by J.~Bardeen, L.~N.~Cooper
and J.~R.~Schrieffer (BCS) in 1957 \cite{Bardeen:1957kj}.
Following the success of the BCS theory, the idea of pairing was
carried over to other areas of physics as well. In particular,
pairing now plays an essential role in the nuclear shell model as
the residual interaction between nucleons and successfully
recounts for various properties of atomic nuclei
\cite{BMP,Belyaev}.

In order to investigate the influence of pairing on nuclear
properties, many authors have used exact analytical solutions of
nuclear shell model which are available in some simplified cases.
For example, in Ref. \cite{Kerman1}, Kerman considered pairs of
nucleons coupled to angular momentum zero occupying a single orbit
and introduced the quasi-spin formalism in which these pairs can
be treated within suitable representations of the angular momentum
algebra. He used this formalism to write down the exact energy
eigenstates and to analyze the influence of pairing on the
collective vibrations of nuclei. Quasi-spin formalism can also be
extended to the case of several orbits in which case the
quasi-spin angular momenta corresponding to different orbits
commute with each other and the pairing term has the form of a
coupling between these angular momenta. This observation
establishes a direct link between the fermion pairing models and
the interacting spin models (see Refs.
\cite{Dukelsky:2004re,Balantekin:2007er,Sierra:2001cx} for
reviews). The exact diagonalization of the later model was carried
out by R.~W.~Richardson in 1962 in the case of the
orbit-independent (i.e., reduced) pairing interaction
\cite{Richardson1}. Later, it was clarified that the exact
solvability of the pairing Hamiltonian in the reduced pairing case
can be understood in terms of a set of quantum invariants which
commute with one another and also with the Hamiltonian
\cite{Gaudin1,Cambiaggio:1997vz}. These invariants are called
rational Gaudin magnet Hamiltonians since they stem from the work
of M. Gaudin who was originally trying to find the largest set
mutually commuting operators for a given system of interacting
spins. In Ref. \cite{Gaudin1}, Gaudin also showed that the pair
creation and annihilation operators which are used in building the
simultaneous eigenstates of the rational Gaudin magnet
Hamiltonians form an algebra which is today known as the rational
Gaudin algebra. It is worth mentioning that the rational Gaudin
algebra is related to the rational solution of the classical
Yang-Baxter equation which appears as an integrability condition
in many contexts. As a result, the rational Gaudin magnet
operators and the rational Gaudin algebra have found many other
applications in physics (see Refs.
\cite{Dukelsky:2004re,Balantekin:2007er,Sierra:2001cx} for reviews
and Refs.
\cite{Lerma:2006ph,Dukelsky:2004aj,Dukelsky:2004uz,Kholodenko:2008yu,Balantekin:2005ks}
for some interesting applications). They have also been
generalized to include other underlying algebraic structures
(i.e., higher rank algebras, super-algebras and deformed algebras)
besides the angular momentum algebra. There is an extensive
literature on this subject and the interested reader may find the
Refs.
\cite{Gaudin2,Sklyanin:1987ih,Sklyanin:1988yz,Jurco:1989fc,Hikami:1992np,Feigin:1994in,Frenkel:2004qy}
useful.

Although the Richardson-Gaudin solution is successfully used in
nuclear physics, the assumption of reduced pairing sometimes
proves to be too stringent. In many cases, the effective residual
interactions between the nucleons are best described by a pairing
force whose strength differs between the orbits\footnote{See, for
example, the treatment of the neutron pairs occupying the valance
shell of Ni isotopes by Auerbach \cite{Auerbach}.}. The
Hamiltonian
\begin{equation}\label{Hamiltonian 1}
\hat{H}=\sum_{j}\sum_m\varepsilon _{j}\> a_{j \> m}^\dagger a_{j
\> m} -|G|\sum_{j,j^{\prime }}\sum_{m,m^\prime}
c_{j}c_{\>j^{\prime }}\> a_{j \> m}^\dagger a_{j \> -m}^\dagger
a_{j^\prime \> -m^\prime}^{}a_{j^\prime \> m^\prime}^{}
\end{equation}
is frequently used to describe such an orbit dependent pairing
interaction. Here, $j$ denotes the total angular momentum of an
orbit and $\varepsilon_j$ denotes its energy. The overall strength
of the pairing term against the kinetic term is measured by the
constant $|G|$ which has the dimension of energy whereas the
relative pairing strengths are measured by the dimensionless
constants $c_{j}$. Richardson-Gaudin scheme mentioned above
applies to the special case of this Hamiltonian in which all
$c_j$'s are the same whereas all single particle energy levels
$\varepsilon_j$ are different from one another (the reduced
pairing case). The focus of this paper, however, is the opposite
case in which all the $c_j$'s are different from one another and
all single particle energies $\varepsilon_j$ are the same (the
degenerate case).

The problem described by  Hamiltonian given in Eq.
(\ref{Hamiltonian 1}) is exactly solvable in both the reduced
pairing case and the degenerate case. As mentioned above, in the
reduced pairing case the solution was given by Richardson-Gaudin
scheme and the corresponding quantum invariants are the rational
Gaudin magnet Hamiltonians. In the degenerate case, the exact
energy eigenvalues and eigenstates were obtained in a series of
papers by Pan \textit{et al} \cite{Pan:1997rw} and by Balantekin
\textit{et al} \cite{Balantekin:2007vs,Balantekin:2007qr} and the
purpose of the present paper is to identify the corresponding
quantum invariants in the degenerate case. In addition, it will be
shown that the quantum invariants in the degenerate case can be
simultaneously diagonalized with the help of the algebraic Bethe
ansatz method. An interesting observation regarding the Bethe
ansatz diagonalization is that although the reduced pairing and
the degenerate cases are opposite of each other, the Bethe ansatz
diagonalization of the invariant operators in both cases is
connected with the rational Gaudin algebra.

The organization of this paper is as follows: Section II is a
brief review of the quasispin formalism and it also serves to
introduce some notation. In Section III, a short review of the
Richardson-Gaudin formalism and the rational Gaudin magnet
Hamiltonians is presented. The main results of this paper, i.e.,
the quantum invariants in the degenerate case and their
simultaneous diagonalization with the Bethe ansatz method are
presented in Section IV. This section also contains a discussion
about a symmetry in the spectra of these quantum invariants
relating eigenvalues corresponding to different number of
particles. In Section V, we consider the rational Gaudin algebra
and point out its relationship with the Bethe ansatz
diagonalization in both the reduced pairing and the degenerate
cases. Section VI summarizes the main conclusions. The details of
some of the Bethe ansatz calculations can be found in the
Appendix.

\section{Quasi-Spin Formalism and the Exact Solutions of the Pairing Hamiltonian}

In the quasi-spin formalism, nucleon pairs coupled to angular
momentum zero are created and annihilated at the level $j$ by the
operators
\begin{equation}\label{Quasispin Operators +-}
S_{j}^+=\sum_{m>0}(-1)^{j-m}a_{j \> m}^\dagger a_{j \> -m}^\dagger
\quad\quad S_{j}^{-}=\sum_{m>0}(-1)^{j-m}a_{j \> -m}a_{j \> m},
\end{equation}
respectively. Together with the operator
\begin{equation}\label{Quasispin Operators 0}
S_j^0=\frac{1}{2}\sum_{m>0}\left(a_{j \> m}^\dagger a_{j \> m}+
a_{j \> -m}^\dagger a_{j \> -m} -1\right)
\end{equation}
they obey the well known angular momentum commutations relations
\begin{equation}\label{Quasispin Algebra}
[S_j^+,S_{j^\prime}^-]=2\delta_{jj^\prime}S_j^0 \quad\quad
[S_j^0,S_{j^\prime}^\pm]=\pm\delta_{jj^\prime}S_j^\pm.
\end{equation}
As a result, one has an angular momentum algebra (the so called
quasi-spin algebra) for each orbit $j$ such that those angular
momenta corresponding to different orbits commute with one
another. The pair number operator for the orbit $j$ is given by
\begin{equation}
\hat{N}_j=\frac{1}{2}\sum_{m>0}\left(a_{j \> m}^\dagger a_{j \>
m}+ a_{j \> -m}^\dagger a_{j \> -m} \right).
\end{equation}
It is related to the operator $S_j^0$ given in Eq. (\ref{Quasispin
Operators 0}) by the formula
\begin{equation}\label{Quasispin Operators 0 Alternative}
S_j^0=\hat{N}_j-\frac{\Omega_j}{2}
\end{equation}
where $\Omega_j$ is the maximum number of pairs which can occupy
the level $j$. Note that $j$ is always an half integer because of
the spin-orbit coupling in the nuclei. As a result, if there are
no unpaired particles at the level $j$, then
\begin{equation}
\Omega_j=j+\frac{1}{2}.
\end{equation}
Also note that the pairing term in the Hamiltonian given in Eq.
(\ref{Hamiltonian 1}) does not act on the unpaired particles. If
there is an unpaired particle at the level $j$, its effect will be
i) to add a constant $\varepsilon_j$ to the Hamiltonian because of
the kinetic term and ii) to reduce the maximum number of pairs
which can occupy the level $j$ by one, i.e., to take $\Omega_j$ to
$\Omega_j-1$. But here it will be assumed that there are no
unpaired particles in the system. In this case, Eq.
(\ref{Quasispin Operators 0 Alternative}) implies that
\begin{equation}
-\frac{\Omega_j}{2} \leq S_j^0 \leq \frac{\Omega_j}{2}
\end{equation}
i.e., quasi-spin algebra corresponding to the level $j$ is
realized in the $\Omega_j/2$ representation. Therefore, in
addition to the physical angular momentum quantum number $j$, we
also have the \textit{quasi-spin quantum number} $\Omega_j/2$ for
each level. The states
\begin{equation}
|\frac{\Omega_j}{2}, \> -\frac{\Omega_j}{2}\rangle
\quad\quad\mbox{and}\quad\quad |\frac{\Omega_j}{2}, \>
\frac{\Omega_j}{2}\rangle
\end{equation}
respectively represent the situations in which i) the level $j$ is
not occupied by any pairs and ii) it is maximally occupied by
pairs. In the presence of several orbits with angular momenta
$j_1, j_2, \dots, j_n$, the state
\begin{equation}\label{Empty Shell}
|0\rangle=|\frac{\Omega_{j_1}}{2}, \>
-\frac{\Omega_{j_1}}{2}\rangle
\otimes %
|\frac{\Omega_{j_2}}{2}, \> -\frac{\Omega_{j_2}}{2}\rangle%
\otimes \dots \otimes %
|\frac{\Omega_{j_n}}{2}, \> -\frac{\Omega_{j_n}}{2}\rangle
\end{equation}
represents a shell which contains no pairs whereas the state
\begin{equation}\label{Full Shell}
|\bar{0}\rangle=|\frac{\Omega_{j_1}}{2}, \>
\frac{\Omega_{j_1}}{2}\rangle
\otimes %
|\frac{\Omega_{j_2}}{2}, \> \frac{\Omega_{j_2}}{2}\rangle%
\otimes \dots \otimes %
|\frac{\Omega_{j_n}}{2}, \> \frac{\Omega_{j_n}}{2}\rangle
\end{equation}
represents a shell which is fully occupied by pairs.

The pairing Hamiltonian given in Eq. (\ref{Hamiltonian 1}) can be
written in terms of the quasi-spin operators given in Eqs.
(\ref{Quasispin Operators +-}) and (\ref{Quasispin Operators 0})
as
\begin{equation}\label{Hamiltonian 2}                                
\hat{H}=\sum_{j}\varepsilon _{j}\left( 2S_{j}^{0}+\Omega
_{j}\right)
-|G|(\sum_j c_{j}S_{j}^{+})%
(\sum_{j^{\prime}}c_{j^{\prime}} S_{j^{\prime}}^-).
\end{equation}
Note that the operator $\sum_j c_{j}S_{j}^{+}$ in Hamiltonian
(\ref{Hamiltonian 2}) creates a pair of particles in such a way
that $c_j$ can be viewed as the probability amplitude that this
pair is found at the level $j$. For this reason the coefficients
$c_j$ are usually called \textit{occupation probability
amplitudes} and they are normalized as
\begin{equation}\label{Normalization}
\sum_j c_j^2=1.
\end{equation}
Although an occupation probability amplitude is a complex number
in general, the parameters $c_j$ can be taken as real without loss
of generality. Because if one $c_j$ is complex, a unitary
transformation can always be performed on the quasi-spin algebra
corresponding to the level $j$ to make that $c_j$ real. Also note
that the Hamiltonian in Eq. (\ref{Hamiltonian 2}) contains a
constant term $\sum_j 2\varepsilon_j\Omega_j$ which comes from Eq.
(\ref{Quasispin Operators 0 Alternative}). This constant term is
not dropped because it guarantees that the energy of the empty
shell is zero, i.e.,
\begin{equation}
\hat{H}|0\rangle=0.
\end{equation}
Using the commutators given in Eq. (\ref{Quasispin Algebra}), one
can show that the fully occupied shell $|\bar{0}\rangle$ is also
an eigenstate of the Hamiltonian with the energy
\begin{equation}
\hat{H}\>|\bar{0}\rangle=\sum_j\left(2\varepsilon_j-|G|c_j^2\right)
\Omega_j\>|\bar{0}\rangle.
\end{equation}
Unlike the empty shell $|0\rangle$ and the fully occupied shell
$|\bar{0}\rangle$, the eigenstates of the pairing Hamiltonian
corresponding to a partially occupied valance shell are unknown in
the most general case. But, as mentioned in the Introduction,
exact energies and eigenstates are known in the two opposite
cases. Namely, the reduced pairing case characterized by
\begin{eqnarray}
\varepsilon_1& \neq &\varepsilon_2 \neq \dots \neq \varepsilon_n \nonumber  \\
c_1&=&c_2=\dots=c_n, \label{Condition1}
\end{eqnarray}
and the degenerate case characterized by
\begin{eqnarray}
\varepsilon_1&=&\varepsilon_2=\dots=\varepsilon_n  \nonumber \\
c_1& \neq &c_2 \neq \dots \neq c_n. \label{Condition2}
\end{eqnarray}
These solutions will be reviewed in the next two sections together
with the corresponding quantum invariants. But before closing this
section, mention must be made of a third case in which exact
eigenstates of the pairing Hamiltonian are known. This solution is
available in the presence of two orbits with unequal energies and
unequal occupation probability amplitudes, i.e.,
\cite{Balantekin:2007ip}:
\begin{eqnarray}
\varepsilon_1& \neq &\varepsilon_2\nonumber \\
c_1& \neq &c_2 . \label{Condition3}
\end{eqnarray}
But this third case will not be considered in this paper. Because
the main interest of this paper is the quantum invariants of the
pairing Hamiltonian and in the case of a two level system we have
only two quantum invariants which are simply the Hamiltonian
itself and the total pair number operator.

\section{Reduced Pairing and the Gaudin Magnet Operators}

In the reduced pairing case, described by Eq. (\ref{Condition1}),
the pairing Hamiltonian given in Eq. (\ref{Hamiltonian 2}) becomes
\begin{equation}\label{Hamiltonian 3}                                
\hat{H}_R=\sum_{j}\varepsilon _{j}\left( 2S_{j}^{0}+\Omega
_{j}\right) -|G|d \sum_{j,j^\prime}S_j^{+}S_{j^\prime}^-.
\end{equation}
Here $d=1/n$ is known as the level spacing and its appearance in
the Hamiltonian is due to the normalization condition
(\ref{Normalization}). Using a variational technique, Richardson
showed in Ref. \cite{Richardson1} that the eigenstates of the
Hamiltonian given in Eq. (\ref{Hamiltonian 3}) containing $N$
pairs of particles are in the form
\begin{equation}\label{Richardson Eigenstates}
J^+(\xi_1)J^+(\xi_2)\dots J^+(\xi_N)|0\rangle
\end{equation}
where the pair creation operators $J^+(\xi)$ are given by
\begin{equation}\label{Gaudin Operators}
J^+(\xi)=\sum_j \frac{S_j^+}{2\varepsilon_j-\xi}
\end{equation}
and $|0\rangle$ is the state with no pairs defined in Eq.
(\ref{Empty Shell}). The values of the parameters $\xi_1, \xi_2,
\dots, \xi_N$ which appear in Eq. (\ref{Richardson Eigenstates})
are to be determined by solving the system of equations
\begin{equation}\label{Richardson BAEs}
\sum_j\frac{-\Omega_j/2}{2\varepsilon_j-\xi_k}=-\frac{1}{2|G|d}+\sum_{\substack{l=1
\\ l\neq k}}^N\frac{1}{\xi_k-\xi_l}
\end{equation}
simultaneously for $k=1,2,\dots,N$ (see Ref. \cite{Richardson1}).
These equations generally have several distinct solutions. For
each one of these solutions we have an eigenstate in the form of
Eq. (\ref{Richardson Eigenstates}) and the corresponding energy is
given by
\begin{equation}\label{Richardson Energy}
E^{(N)}=\sum_{k=1}^N \xi_k .
\end{equation}

The quantum invariants of the Hamiltonian in Eq. (\ref{Hamiltonian
3}) are the rational Gaudin magnet Hamiltonians mentioned in the
introduction \cite{Gaudin1,Cambiaggio:1997vz}. They are given
by\footnote{Strictly speaking, the operators studied by Gaudin
himself did not include the one body term $S_j^0$.}
\begin{equation}\label{R}
\hat{R}_j=S^0_j-|G|d\sum_{j^\prime(\neq j)}%
\frac{\overrightarrow{S}_j\cdot\overrightarrow{S}_{j^\prime}}{\varepsilon_j-\varepsilon_{j^\prime}}
\end{equation}
where $\overrightarrow{S}_j\cdot\overrightarrow{S}_{j^\prime}$ is
defined as
\begin{equation}
\overrightarrow{S}_j\cdot\overrightarrow{S}_{j^\prime}=%
S_j^0 S_{j^\prime}^0%
+\frac{1}{2}\left(S_j^+ S_{j^\prime}^-%
+S_j^- S_{j^\prime}^+\right).
\end{equation}
The rational Gaudin magnet Hamiltonians mutually commute with one
another and with the Hamiltonian $\hat{H}_R$, i.e.,
\begin{equation}\label{Integrals of Motion}
[\hat{R}_j,\hat{R}_{j^\prime}]=0 \quad\quad\quad
[\hat{R}_j,\hat{H}_R]=0
\end{equation}
for every $j,j^\prime=1,2,\dots,n$. The Hamiltonian itself is not
an independent invariant and it can be written in terms of the
operators $\hat{R}_j$ as
\begin{equation}\label{Hamiltonian 4}
\hat{H}_R=\sum_j \left(2\varepsilon_j-|G|d\right) \hat{R}_j %
+|G|d\sum_{j,j^\prime} \hat{R}_j \hat{R}_{j^\prime}%
-|G|d\sum_j
\overrightarrow{S}_j\cdot\overrightarrow{S}_{j} %
+\sum_j \varepsilon_j \Omega_j.
\end{equation}
Similarly, the total pair number operator can also be written in
terms of the operators $\hat{R}_j$ as
\begin{equation}
\hat{N}=\sum_j\left(\hat{R}_j + \frac{\Omega_j}{2}\right)
\end{equation}

As a result of Eq. (\ref{Integrals of Motion}),  the eigenstates
of the Hamiltonian are at the same time simultaneous eigenstates
of rational Gaudin magnet operators as well. Let us denote the
eigenvalues of the invariant $\hat{R}_j$ corresponding to the
eigenstate with $N$ pairs by $E_j^{(N)}$. In other words,
\begin{equation}\label{Ricardson_Eigenvalue_0}
\hat{R}_j|0\rangle=E_j^{(0)}|0\rangle
\end{equation}
for the empty shell and
\begin{equation}\label{Ricardson_Eigenvalue_N}
\hat{R}_j\> J^+(\xi_1)J^+(\xi_2)\dots
J^+(\xi_N)|0\rangle=E_j^{(N)}\> J^+(\xi_1)J^+(\xi_2)\dots
J^+(\xi_N)|0\rangle
\end{equation}
for the eigenstates containing $N$ pairs described by Eqs.
(\ref{Richardson Eigenstates}-\ref{Richardson BAEs}). The
eigenvalues $E_j^{(0)}$ and $E_j^{(N)}$ are given by
\begin{equation}\label{E_j^0}
E_j^{(0)}=-\frac{\Omega_j}{2}-\frac{|G|d}{4}\sum_{j(\neq
j^\prime)}\frac{\Omega_j\Omega_{j^\prime}}{\varepsilon_j-\varepsilon_{j^\prime}}.
\end{equation}
and
\begin{equation}\label{E_j^N}
E_j^{(N)}=E_j^{(0)}+|G|d\sum_{k=1}^N
\frac{\Omega_j}{2\varepsilon_j-\xi_k}
\end{equation}
respectively.

The pairing Hamiltonian in Eq. (\ref{Hamiltonian 3}) is only one
of the exactly solvable models which can be built using the
rational Gaudin magnet Hamiltonians. Various other linear or
nonlinear combinations of rational Gaudin magnet operators can be
used to built other useful exactly solvable models (see, for
example, Refs.
\cite{Dukelsky:2004aj,Dukelsky:2004uz,Balantekin:2005ks}).

\section{Integrability in the Degenerate Case}

In the case of several orbits having the same energy but different
occupation probability amplitudes, i.e., when the conditions in
Eq. (\ref{Condition2}) are satisfied, the first term in the
pairing Hamiltonian given in Eq. (\ref{Hamiltonian 2}) becomes a
constant which is proportional to the total number of pairs in the
shell. Discarding this term, one can write the Hamiltonian as
\begin{equation}\label{Hamiltonian 5}
\hat{H}_D=-|G|(\sum_j c_{j}S_{j}^{+})%
(\sum_{j^{\prime}}c_{j^{\prime}} S_{j^{\prime}}^-).
\end{equation}
Exact eigenvalues and eigenstates of the Hamiltonian given in Eq.
(\ref{Hamiltonian 5}) were obtained in Refs.
\cite{Pan:1997rw,Balantekin:2007vs,Balantekin:2007qr}. The purpose
of this paper is to introduce the corresponding quantum
invariants, i.e., the set of operators which commute with one
another and with the Hamiltonian in Eq. (\ref{Hamiltonian 5}).

The fact that the rational Gaudin magnet operators given in Eqs.
(\ref{R}) mutually commute with one another is independent of the
values of the parameters $\varepsilon_j$. Naturally, one can try
to replace the parameters $\varepsilon_j$ in the Gaudin operators
with some arbitrary functions of $c_j$ and try to determine the
form of these functions so that the new operators commute with the
Hamiltonian in Eq. (\ref{Hamiltonian 5}) as well. It turns out,
however, that such as course of action does not yield the quantum
invariants of the  Hamiltonian given in Eq. (\ref{Hamiltonian
5})\footnote{In addition to the rational Gaudin magnet
Hamiltonians, Gaudin also studied the so called trigonometric and
hyperbolic Gaudin magnet Hamiltonians which also mutually commute
with one another. Similar to the rational case, however, one
cannot choose the parameters of trigonometric or hyperbolic Gaudin
magnet Hamiltonians so as to make them commute with the
Hamiltonian in Eq. (\ref{Hamiltonian 5}).}.

In order to find the invariant operators one can consider general
number conserving operators in the form
\begin{equation}
\hat{P}_{j}=A_{j}S_{j}^{0}+B_{j}S_{j}^{+}S_{j}^{-}+\sum_{j^{\prime
}\left( \not=j\right) }D_{jj^{\prime }}S_{j}^{0}S_{j^{\prime
}}^{0}+\sum_{j^{\prime }\left( \not=j\right) }F_{jj^{\prime
}}\left( S_{j}^{+}S_{j^{\prime }}^{-}+S_{j^{\prime
}}^{+}S_{j}^{-}\right)
\end{equation}
where $A_j$, $B_j$ $D_{jj^{\prime }}$ and $F_{jj^{\prime }}$ are
some arbitrary coefficients. The condition that the above
operators commute with one another and with the Hamiltonian in Eq.
(\ref{Hamiltonian 5}) gives us the allowed values of these
coefficients. A straightforward calculation shows that the desired
operators are given by
\begin{equation}\label{Constants}
\hat{P}_{j}=-S_{j}^{+}S_{j}^{-}+2\sum_{{j^\prime}\left(\neq
j\right)
}\frac{c_{{j^\prime}}^{2}}{c_{{j^\prime}}^{2}-c_{j}^{2}}S_{j}^{0}S_{{j^\prime}}^{0}
+\sum_{j^\prime \\ \left(\neq
j\right)}\frac{c_{j}c_{{j^\prime}}}{c_{{j^\prime}}^{2}-c_{j}^{2}}\left(
S_{j}^{+}S_{{j^\prime}}^{-}+S_{{j^\prime}}^{+}S_{j}^{-}\right).
\end{equation}
These operators mutually commute with one another
\begin{equation}\label{mutual commutation}
\left[\hat{P}_{j},\hat{P}_{j^\prime}\right]=0
\end{equation}
for every $j$ and $j^\prime$. They also commute with the
Hamiltonian given in Eq. (\ref{Hamiltonian 5}) and with the total
pair number operator:
\begin{equation}\label{commutation with H}
\left[\hat{P}_{j},\hat{H}_D\right]=0 \quad\quad\quad\left[
\hat{P}_{j},\hat{N}\right]=0.
\end{equation}
The Hamiltonian $\hat{H}_D$ and the total number operator
$\hat{N}$ are not independent invariants but they are related to
the operators $\hat{P}_j$ by the formulas
\begin{equation}\label{Hamiltonian 6}
|G|\sum_{j}c_{j}^{2}\hat{P}_{j}=\hat{H}_D
\end{equation}
and
\begin{equation}
\sum_{j}\hat{P}_{j}=\widehat{N}^{2}-\widehat{N}(\sum_{j}\Omega_{j}+1)
+\frac{1}{4}\sum_{\substack{j,j^\prime
\\ \left( j\not=j^\prime \right) }}\Omega _{j}\Omega _{j^{\prime
}}.
\end{equation}

As a result of Eqs. (\ref{mutual commutation}) and
(\ref{commutation with H}), the invariants $\hat{P}_{j}$ have the
same eigenstates as the pairing Hamiltonian $\hat{H}_D$ given in
Eq. (\ref{Hamiltonian 5}). These eigenstates were given in Refs.
\cite{Pan:1997rw,Balantekin:2007vs,Balantekin:2007qr} with the
help of the Bethe ansatz method \cite{Bethe:1931hc}. In what
follows, the corresponding eigenvalues of the invariant operators
$\hat{P}_{j}$ will be presented. A summary of the results of this
Section can be found in Table \ref{Table 1}.

Following Refs.
\cite{Pan:1997rw,Balantekin:2007vs,Balantekin:2007qr}, let us
introduce the pair creation and annihilation operators
\begin{equation}\label{Define S}
S^+(x)=\sum_j\frac{c_j}{1-c_j^2 x}S_j^+ \quad\quad
S^-(x)=\sum_j\frac{c_j}{1-c_j^2 x}S_j^-.
\end{equation}
Here, $x$ is a complex variable and $S_j^\pm$ are the quasispin
operators introduced in Eq. (\ref{Quasispin Operators +-}). The
Hamiltonian in Eq. (\ref{Hamiltonian 5}) itself can be written in
terms of these operators as
\begin{equation}\label{Hamiltonian 7}
\hat{H}_D=-|G|\hat{S}^+(0)\hat{S}^-(0).
\end{equation}

The eigenstates of the Hamiltonian $\hat{H}_D$ which are also
simultaneous eigenstates of the invariants $\hat{P}_j$ can be
written in terms of the pair creation and annihilation operators
in Eq. (\ref{Define S}). Below, these eigenstates which were
obtained in Refs.
\cite{Pan:1997rw,Balantekin:2007vs,Balantekin:2007qr} will be
reviewed in the order of increasing number of pairs and the
corresponding eigenvalues of the invariant operators $\hat{P}_j$
will be given.

\vskip 2mm

{\bf Empty shell:} The empty shell $|0\rangle$ given in Eq.
(\ref{Empty Shell}) obeys
\begin{equation}
\hat{P}_j|0\rangle=E_j^{(0)}|0\rangle
\end{equation}
where $E_j^{(0)}$ is given by
\begin{equation}\label{E^0_j}
E_j^{(0)}=\frac{\Omega_j}2\sum_{j^\prime (\neq j)}
\frac{\Omega_{j^\prime }}{1-c_j^2/c_{j^\prime }^2}.
\end{equation}

\vskip 2mm

{\bf Eigenstates with $N=1$:} The eigenstates with one pair of
particles fall in two classes. The state
\begin{equation}\label{Eigenstate N=1 0}
\hat{S}^+(0)|0\rangle
\end{equation}
is an eigenstate where $\hat{S}^+(0)$ is obtained by putting $x=0$
in the operator given in Eq. (\ref{Define S}). This state was
first suggested by Talmi in Ref. \cite{Talmi1} and was shown to be
an eigenstate of a class of Hamiltonians including the Hamiltonian
in Eq. (\ref{Hamiltonian 5}).

In addition to the state in Eq. (\ref{Eigenstate N=1 0}), the
state
\begin{equation}\label{Eigenstate N=1 x}
\hat{S}^+(x)|0\rangle
\end{equation}
is also an eigenstate if $x$ is a solution of the \Bae
\begin{equation}\label{BAE N=1 x}
\sum_j \frac{-\Omega_j/2}{1/c_j^2-x}=0.
\end{equation}

The eigenvalues of the operators $\hat{P}_j$ corresponding to the
eigenstates described above will be denoted by $\lambda_j^{(1)}$
and $\mu_j^{(1)}$, respectively, i.e.,
\begin{eqnarray}
\hat{P}_j\>\hat{S}^+(0)|0\rangle
&=&\lambda_j^{(1)}\>\hat{S}^+(0)|0\rangle \label{A1}\\
\hat{P}_j\>\hat{S}^+(x)|0\rangle
&=&\mu_j^{(1)}\>\hat{S}^+(x)|0\rangle\label{A2}.
\end{eqnarray}
The eigenvalues $\lambda_j^{(1)}$ and $\mu_j^{(1)}$ can easily be
computed using the commutators given in Eq. (\ref{Quasispin
Algebra}) together with Eqs. (\ref{Define S}) and (\ref{BAE N=1
x}) as follows (see the Appendix)
\begin{eqnarray}
\lambda_j^{(1)}&=&E_j^{(0)}-\Omega_j \label{lambda_j^1}\\
\mu_j^{(1)} &=&
E_j^{(0)}-\frac{\Omega_j}{1-c_j^{2}x}\label{mu_j^1}.
\end{eqnarray}
Note that the eigenstate in Eq. (\ref{Eigenstate N=1 0}) is unique
whereas the eigenstate in Eq. (\ref{Eigenstate N=1 x}) represents
several eigenstates. Because in general the \Bae ~(\ref{BAE N=1
x}) has more than one solutions and for each one of them we have
an eigenstate in the form of Eq. (\ref{Eigenstate N=1 x}). As a
result, Eq. (\ref{A2}) also represents several
eigenvalue-eigenstate equations.

\vskip 2mm

{\bf Eigenstates for $2\leq N\leq N_{max}/2$:} The results given
above can be generalized to the states corresponding to a shell
which is at most half full. Let $N_{max}=\sum_j\Omega_j$ denote
the maximum number of pairs which can occupy the shell in
consideration. Then for $2\leq N\leq N_{max}/2$ the results
obtained for one pair generalizes as follows: The state
\begin{equation}\label{Eigenstate N 0}
\hat{S}^+(0)\hat{S}^+(z_1) \dots \hat{S}^+(z_{N-1})|0\rangle
\end{equation}
which has $N$ pairs of particles is an eigenstate if the
parameters $z_k$ are all different from one another and obey the
following system of Bethe ansatz equations
\begin{equation}\label{BAE N 0}
\sum_j \frac{-\Omega_j/2}{1/c_j^2-z_m}
=\frac{1}{z_m}+\sum_{\substack {k=1\\ (k\neq m)}}^{N-1}
\frac{1}{z_m-z_k},
\end{equation}
for every $m=1,2,\dots N-1$. In addition, the state
\begin{equation}\label{Eigenstate N x}
\hat{S}^+(x_1)\hat{S}^+(x_2) \dots \hat{S}^+(x_N)|0\rangle
\end{equation}
which also has $N$ pairs of particles is an eigenstate if the
parameters $x_k$ are all different from one another and satisfy
the following system of Bethe ansatz equations:
\begin{equation}\label{BAE N x}
\sum_j \frac{-\Omega_j/2}{1/|c_j|^2-x_m}=\sum_{\substack {k=1\\
(k\neq m)}}^N \frac{1}{x_m-x_k},
\end{equation}
for every $m=1,2,\dots,N$. Note that the states given in Eqs.
(\ref{Eigenstate N 0}) and (\ref{Eigenstate N x}) can be thought
of as the generalizations of the states in Eqs. (\ref{Eigenstate
N=1 0}) and (\ref{Eigenstate N=1 x}), respectively.

The \Baes ~given in Eqs. (\ref{BAE N 0}) and (\ref{BAE N x}) have
more than one solutions.\footnote{It is worth emphasizing that in
driving Eqs. (\ref{BAE N 0}), all the parameters $z_k$ were
assumed to be different from each another and similarly for the
Eqs. (\ref{BAE N x}). As a result, any solution which contradicts
this assumption should be discarded. In fact, it can be shown that
if any two parameters in the state (\ref{Eigenstate N 0}) are
equal to each other, then this state cannot be an eigenstate. A
similar result is also valid for the state in Eq. (\ref{Eigenstate
N x}).} Each one of these solutions gives us an eigenstate in the
form of Eqs. (\ref{Eigenstate N 0}) and (\ref{Eigenstate N x}),
respectively. One should keep in mind, however, that the states in
Eqs. (\ref{Eigenstate N 0}) and (\ref{Eigenstate N x}) are
invariant under the permutations of the parameters and so are the
corresponding \Baes. Consequently, solutions of the \Baes ~which
differ only by a reordering of the variables should be counted as
one solution.

The eigenvalues of the operators $\hat{P}_j$ corresponding to the
eigenstates given in Eqs. (\ref{Eigenstate N 0}) and
(\ref{Eigenstate N x}) above will be denoted by $\lambda_j^{(N)}$
and $\mu_j^{(N)}$, respectively, i.e.,
\begin{eqnarray}
\hat{P}_j\> \hat{S}^+(0)\hat{S}^+(z_1) \dots
\hat{S}^+(z_{N-1})|0\rangle &=&\lambda_j^{(N)}\>
\hat{S}^+(0)\hat{S}^+(z_1) \dots \hat{S}^+(z_{N-1})|0\rangle \label{A3}\\
\hat{P}_j\>\hat{S}^+(x_1)\hat{S}^+(x_2) \dots
\hat{S}^+(x_N)|0\rangle&=&\mu_j^{(N)}\>
\hat{S}^+(x_1)\hat{S}^+(x_2) \dots
\hat{S}^+(x_N)|0\rangle\label{A4}
\end{eqnarray}
These eigenvalues can be computed with the help of Eqs.
(\ref{Quasispin Algebra}), (\ref{Define S}), (\ref{BAE N 0}) and
(\ref{BAE N x}). The results are given by (see the Appendix for
the details)
\begin{eqnarray}
\lambda^{(N)}_j&=&E_j^{(0)}-\Omega_j-\sum_{k=1}^{N-1}\frac{\Omega_j}{1-c_j^2z_k} \label{lambda_j^N},\\
\mu^{(N)}_j&=&E_j^{(0)}-\sum_{k=1}^N\frac{\Omega_j}{1-c_j^2
x_k}\label{mu_j^N}
\end{eqnarray}


\begin{sidewaystable}[ht]
\caption{Summary of the energy eigenvalues and the eigenstates of
the pairing Hamiltonian in the degenerate limit, given in Eq.
(\ref{Hamiltonian 5}) and its quantum invariants given in Eq.
(\ref{Constants}). Here, $N_{max}$ denotes the maximum number of
pairs which can occupy the shell.}\label{Table 1}
\begin{tabular}{llll}
\hline \multicolumn{4}{l}{\textbf{Empty shell: }${\displaystyle
\mathbf{N=0}}$}
\\ \hline & & &  
\\ State & Bethe Ansatz Equation & Eigenvalue of
${\displaystyle \widehat{P}_{j}}$ & Eigenvalue of ${\displaystyle
\widehat{H}_{D}}$
\\ & & & 
\\ ${\displaystyle |0\rangle }$ & No BAE & ${\displaystyle E_{j}^{\left( 0\right) }=\frac{\Omega _{j}}{2}%
\sum_{\substack {j^{\prime }\\ \left(j^\prime \not=j\right) }}\frac{\Omega _{j^{\prime }}}{%
1-c_{j}^{2}/c_{j^{\prime }}^{2}}}$ & ${\displaystyle 0 }$
\\ \hline \multicolumn{4}{l}{\textbf{One pair of particles in the shell:
}${\displaystyle \mathbf{N=1} }$}
\\ \hline & & & 
\\ State & Bethe Ansatz Equation  & Eigenvalue of ${\displaystyle \widehat{P}_{j}}$ &
Eigenvalue of ${\displaystyle \widehat{H}_{D}}$
\\ & & & 
\\ ${\displaystyle S^{+}\left( x\right) |0\rangle }$ &  ${\displaystyle\sum_{j}\frac{\Omega _{j}}{1/c_{j}^{2}-x%
}=0}$ & ${\displaystyle E_{j}^{\left( 0\right) }-\frac{\Omega
_{j}}{1-c_{j}^{2}x}}$ & ${\displaystyle 0 }$
\\ ${\displaystyle S^{+}\left( 0\right) |0\rangle }$ & No BAE & ${\displaystyle E_{j}^{\left( 0\right)
}-\Omega _{j} }$ & ${\displaystyle -\left\vert G\right\vert
\sum_{j}c_{j}^{2}\Omega _{j}}$
\\\hline\multicolumn{4}{l}{\textbf{At most half full Shell: }${\displaystyle \mathbf{N}\leq \mathbf{%
N}_{\max }\mathbf{/2\ }}$}
\\ \hline & & & 
\\ State & Bethe Ansatz Equation  & Eigenvalue of ${\displaystyle \widehat{P}_{j}}$ &
Eigenvalue of ${\displaystyle \widehat{H}_{D}}$
\\ & & & 
\\ ${\displaystyle S^{+}\left( x_{1}\right) S^{+}\left( x_{2}\right) \ldots
S^{+}\left(
x_{N}\right) |0\rangle }\quad\quad\quad\quad $ & ${\displaystyle \sum_{j}\frac{\Omega _{j}}{1/c_{j}^{2}-x_{k}}%
+\sum_{\substack {l=1 \\ \left( l\not=k\right)} }^{N}\frac{2}{x_{k}-x_{l}}=0}$ & ${\displaystyle %
E_{j}^{\left( 0\right) }-\sum_{k=1}^{N}\frac{\Omega
_{j}}{1-c_{j}^{2}x_{k}}}\quad\quad $ & ${\displaystyle 0}$
\\ ${\displaystyle S^{+}\left( 0\right) S^{+}\left( z_{1}\right) \ldots S^{+}\left(
z_{N-1}\right) |0\rangle }$ & ${\displaystyle \sum_{j}\frac{\Omega _{j}}{1/c_{j}^{2}-z_{k}}+%
\frac{2}{z_{k}}+\sum_{\substack {l=1 \\ \left( l\not=k\right)}
}^{N-1}\frac{2}{z_{k}-z_{l}}=0}\quad\quad\quad$
& ${\displaystyle E_{j}^{\left( 0\right) }-\Omega _{j}-\sum_{k=1}^{N-1}\frac{\Omega _{j}}{%
1-c_{j}^{2}z_{k}}}\quad\quad\quad$ & ${\displaystyle -\left\vert
G\right\vert \sum_{j}c_{j}^{2}\Omega _{j}+\left\vert G\right\vert
\sum_{k=1}^{N-1}\frac{2}{z_{k}}}$
\\ \hline \multicolumn{4}{l}{\textbf{More than half full Shell: }${\displaystyle \mathbf{N}_{\max }%
\mathbf{/2}<\mathbf{N} }$}
\\ \hline & & & 
\\ State & Bethe Ansatz Equation  & Eigenvalue of ${\displaystyle \widehat{P}_{j}}$ &
Eigenvalue of ${\displaystyle \widehat{H}_{D}}$
\\ & & & 
\\ ${\displaystyle S^{-}\left( z_{1}\right) S^{-}\left( z_{2}\right) \ldots
S^{-}\left(
z_{N-1}\right) |\overline{0}\rangle }$ & ${\displaystyle \sum_{j}\frac{\Omega _{j}}{%
1/c_{j}^{2}-z_{k}}+\frac{2}{z_{k}}+\sum_{\substack {l=1 \\ \left( l\not=k\right)} }^{N-1}%
\frac{2}{z_{k}-z_{l}}=0}$ & ${\displaystyle E_{j}^{\left( 0\right)
}-\Omega _{j}-\sum_{k=1}^{N-1}\frac{\Omega
_{j}}{1-c_{j}^{2}z_{k}}}$ & ${\displaystyle -\left\vert
G\right\vert \sum_{j}c_{j}^{2}\Omega _{j}+\left\vert G\right\vert
\sum_{k=1}^{N-1}\frac{2}{z_{k}}}$
\\ \hline \multicolumn{4}{l}{\textbf{Full Shell }${\displaystyle \mathbf{N=N}_{\max}}$}
\\ \hline & & & 
\\ State & Bethe Ansatz Equation  & Eigenvalue of $\widehat{P}_{j}$ &
Eigenvalue of ${\displaystyle \widehat{H}_{D}}$
\\ & & & 
\\ ${\displaystyle |\overline{0}\rangle }$ & No BAE & ${\displaystyle E_{j}^{\left( 0\right) }-\Omega _{j}}$
& ${\displaystyle -\left\vert G\right\vert \sum_{j}c_{j}^{2}\Omega
_{j}}$
\\ \hline
\end{tabular}
\end{sidewaystable}%


\vskip 2mm

{\bf Eigenstates for $N_{max}/2 < N$:} In order to write down the
eigenstates and eigenvalues corresponding to a shell which is more
than half full, we introduce the operator
\begin{equation}\label{T}
\hat{T}=\exp{ \left(-i\pi\sum_j
\frac{S^+_j+S^-_j}{2} \right)}
\end{equation}
following Ref. \cite{Balantekin:2007qr}. This operator transforms
the empty shell given in Eq. (\ref{Empty Shell}) into the fully
occupied shell given in Eq. (\ref{Full Shell}), i.e.,
\begin{equation}\label{Transformation of State}
\hat{T}^\dagger |0\rangle=|\bar{0}\rangle.
\end{equation}
In addition, it also transforms the pair creation operators into
pair annihilation operators and visa versa:
\begin{equation}\label{Trasformation}
\hat{T}^\dagger S_j^\pm \hat{T} = S_j^\mp
\quad\quad\mbox\quad\quad\quad \hat{T}^\dagger S_j^0 \hat{T} =-
S_j^0
\end{equation}
Therefore, the operator $\hat{T}^\dagger$ transforms a state
containing $N$ particle pairs into states containing $N$
hole-pairs. However, there is no particle-hole symmetry in the
problem described by the invariants $\hat{P}_j$ as can be easily
verified by showing that
\begin{equation}
\left[\hat{P}_{j},\hat{T}\right]\neq 0.
\end{equation}
On the other hand, it is easy to show that the operator
\begin{equation}\label{Susy Generator}
\hat{B}=\hat{T}^\dagger S^-(0)
\end{equation}
commutes with the invariants $\hat{P}_j$:
\begin{equation}\label{Symmetry}
[\hat{P}_j, \hat{B}]=0 \ \ \ \ \ [\hat{H}_D, \hat{B}]=0.
\end{equation}
This tells us that if $|\psi\rangle$ is a simultaneous eigenstate
of the operators $\hat{P}_j$ so is $\hat{B}|\psi\rangle$ unless
$|\psi\rangle$ is annihilated by $\hat{B}$. It is easy to see from
the definition of $\hat{B}$ that if the state $|\psi\rangle$ has
$N$ particle-pairs, then the state $\hat{B}|\psi\rangle$ has $N-1$
hole-pairs. Because the operator $\hat{B}$ first annihilates one
pair and then replaces the remaining $N-1$ particle-pairs with
hole-pairs. Consider, for example, the eigenstates with one pair
of particles given in Eqs. (\ref{Eigenstate N=1 0}) and
(\ref{Eigenstate N=1 x}). One can easily show, using Eqs.
(\ref{Hamiltonian 7}), (\ref{Transformation of State}) and
(\ref{Trasformation}) that
\begin{equation}\label{Symmetry N=1}
\hat{B}\>\hat{S}^+(0)|0\rangle \propto |\bar{0}\rangle
\end{equation}
where $|\bar{0}\rangle$ is the state which is maximally occupied
by pairs defined in Eq. (\ref{Full Shell}). Consequently, the
states $\hat{S}^+(0)|0\rangle$ and $|\bar{0}\rangle$ have the same
eigenvalues given by Eqs. (\ref{lambda_j^1}) and (\ref{mu_j^1}).
On the other hand, the state in Eq. (\ref{Eigenstate N=1 x}) which
also has one pair of particles is annihilated by the operator
$\hat{B}$:
\begin{equation}
\hat{B} \hat{S}^+(x)|0\rangle = 0
\end{equation}
This can be easily verified using the commutators (\ref{Quasispin
Algebra}) and the \Bae ~(\ref{BAE N=1 x}). Similarly, if we act on
the eigenstate with $N$ pairs of particles ($2\leq N \leq
N_{max}/2$) given in Eq. (\ref{Eigenstate N 0}), we find that
\begin{equation}
\hat{B}\>\hat{S}^+(0)\hat{S}^+(z_1) \dots
\hat{S}^+(z_{N-1})|0\rangle %
\propto \hat{S}^-(z_1) \dots \hat{S}^-(z_{N-1})|0\rangle
\end{equation}
As a result the state
\begin{equation}
\hat{S}^-(z_1) \dots \hat{S}^-(z_{N-1})|0\rangle
\end{equation}
which has $N-1$ pairs of holes, has the same eigenvalues as the
state in Eq. (\ref{Eigenstate N 0}), i.e., those given in Eqs.
(\ref{lambda_j^N}) and (\ref{mu_j^N}). On the other hand, the
state in Eq. (\ref{Eigenstate N x}) which also has $N$ pairs of
particles ($2\leq N \leq N_{max}/2$) is annihilated by the
operator $\hat{B}$
\begin{equation}
\hat{B}\>\hat{S}^+(x_1)\hat{S}^+(x_2) \dots
\hat{S}^+(x_{N})|0\rangle =0
\end{equation}
as can be verified by using the commutators (\ref{Quasispin
Algebra}) and the \Baes ~given in Eqs. (\ref{BAE N x}).

The eigenstates and the eigenvalues of the invariants $\hat{P}_j$
described in this section are summarized in Table \ref{Table 1}.
This table also contains the corresponding eigenvalues of the
pairing Hamiltonian $\hat{H}_D$ given in Eq. (\ref{Hamiltonian 5})
obtained from the eigenvalues of its invariants by using the
formula (\ref{Hamiltonian 6}). For example, the eigenvalue of
$\hat{H}_D$ corresponding to the state in Eq. (\ref{Eigenstate N=1
0}) is given by
\begin{equation}
|G|\sum_{j}c_{j}^{2}\lambda_j^{(1)},
\end{equation}
etc \dots The energy eigenvalues of the Hamiltonian $\hat{H}_D$
obtained from Eq. (\ref{Hamiltonian 6}) are in agreement with
those obtained earlier in Refs.
\cite{Pan:1997rw,Balantekin:2007vs,Balantekin:2007qr}.

\section{Rational Gaudin Algebra}

Rational Gaudin algebra naturally appears from the pair creation
and annihilation operators used in building the simultaneous
eigenstates of the rational Gaudin magnet Hamiltonians
\cite{Gaudin1,Gaudin2}. Formally, it is defined as an infinite
dimensional algebra whose generators
$J^+(\lambda)$,$J^-(\lambda)$, $J^0(\lambda)$ depend on a complex
variable $\lambda$ and obey the commutation relations
\begin{eqnarray}
[J^+(\lambda),J^-(\mu)]&=&2\frac{J^0(\lambda)-J^0(\mu)}{\lambda-\mu},\nonumber\\
\label{GAUDIN_ALGEBRA}
[J^0(\lambda),J^{\pm}(\mu)]&=&\pm\frac{J^{\pm}(\lambda)-J^{\pm}(\mu)}{\lambda-\mu},
\end{eqnarray}
for $\lambda\neq\mu$. Commutators of $J^+(\lambda), J^-(\lambda)$
and $J^0(\lambda)$ at the same value of the complex parameter are
given by taking the limit $\mu \rightarrow \lambda$. We are
usually interested, however, with the finite dimensional
(unfaithful) realizations of this algebra. For example, it is not
difficult to show that for any set of real parameters $\alpha_j$
which are all different from one another, the operators
\begin{equation}\label{Gaudin Operators2}
J^+(\alpha_j;\lambda)=\sum_j \frac{S_j^+}{\alpha_j-\lambda}
\quad\quad\quad J^-(\alpha_j;\lambda)=\sum_j
\frac{S_j^-}{\alpha_j-\lambda} \quad\quad\quad
J^0(\alpha_j;\lambda)=\sum_j \frac{S_j^0}{\alpha_j-\lambda}
\end{equation}
form a realization of the rational Gaudin algebra given in Eqs.
(\ref{GAUDIN_ALGEBRA}). Note that in Eqs. (\ref{Gaudin
Operators2}) we explicitly stated the dependence of the generators
on the parameters $\alpha_j$. For different values of these
parameters, one obtains different realizations of the rational
Gaudin algebra.

The realization of the rational Gaudin algebra given in Eqs.
(\ref{Gaudin Operators2}) is related to the pairing problem in
both the reduced pairing case and the degenerate case for
different values of the parameters $\alpha_j$. By taking
$\alpha_j=2\varepsilon_j$ in Eq. (\ref{Gaudin Operators2}), one
obtains the operators given in Eq. (\ref{Gaudin Operators}) which
are used in building the eigenstates in the reduced pairing case,
i.e.,
\begin{equation}\label{Gaudin Operators4}
J^+(2\varepsilon_j;\lambda)=\sum_j
\frac{S_j^+}{2\varepsilon_j-\xi} \quad\quad\quad
J^-(2\varepsilon_j;\lambda)=\sum_j
\frac{S_j^-}{2\varepsilon_j-\xi} \quad\quad\quad
J^0(2\varepsilon_j;\lambda)=\sum_j
\frac{S_j^0}{2\varepsilon_j-\xi}
\end{equation}

The pair creation operators given in Eq. (\ref{Define S}) can also
be obtained from a realization of the rational Gaudin algebra
although a change of basis is necessary. By taking
$\alpha_j=1/c_j$ in Eqs. (\ref{Gaudin Operators2}) one obtains
\begin{equation}\label{Gaudin Operators3}
J^+(1/c_j;\lambda)=\sum_j \frac{S_j^+}{1/c_j-\lambda}
\quad\quad\quad J^-(1/c_j;\lambda)=\sum_j
\frac{S_j^-}{1/c_j-\lambda} \quad\quad\quad
J^0(1/c_j;\lambda)=\sum_j \frac{S_j^0}{1/c_j-\lambda}
\end{equation}
and then the operators in Eq. (\ref{Define S}) can be written as a
linear combination of these operators, i.e.,
\begin{equation}\label{New basis}
S^\pm(x)=\frac{J^\pm(1/c_j;\sqrt{x})+J^\pm(1/c_j;-\sqrt{x})}{2}
\end{equation}

Rational Gaudin algebra defined in Eqs. (\ref{GAUDIN_ALGEBRA})
appears in connection with various integrable models as mentioned
in the Introduction. It is often the case that the step operators
which are used in building the eigenstates live in one of its
realizations. Here we see that the integrability of the pairing
Hamiltonian in the reduced pairing and the degenerate pairing
cases can be studied in two different realizations of the rational
Gaudin algebra, namely those given in Eqs. (\ref{Gaudin
Operators4}) and (\ref{Gaudin Operators3}), respectively.

\section{Conclusions}

In this paper, we have obtained the quantum invariants of the
pairing Hamiltonian in the degenerate case and simultaneously
diagonalized them with the help of the algebraic Bethe ansatz
method. Although in order to obtain the eigenvalues one should
first solve a system of Bethe ansatz equations which are nonlinear
and coupled to each other, solving them usually proves to be much
more convenient then a direct numerical diagonalization method.
Exact analytical methods for solving the Bethe ansatz equations
also exist in some simplified cases (see, for example, Refs.
\cite{Balantekin:2005ks,Balantekin:2007vs,Nurtac}). The quantum
invariants obtained in this paper are the counterparts of the well
known rational Gaudin magnet Hamiltonians which play the same role
in the reduced pairing case. It is worth mentioning that since the
quantum invariants are mutually commuting operators, they can be
used to build various other integrable models besides the ones
considered in this paper.

We also pointed out that the integrability of the pairing
Hamiltonian in both the reduced pairing and the degenerate cases
is connected with the rational Gaudin algebra. The generalizations
of the rational Gaudin algebra to different underlying algebraic
systems (such as higher order Lie algebras, quantum algebras or
superalgebras) have been used to study the reduced pairing model
and the related rational Gaudin magnet Hamiltonians in more
general frameworks. The question then naturally arises whether or
not one can do the same generalization for the degenerate pairing
model and the related quantum invariants too.  For example can the
invariant operators of the degenerate pairing given in Eq.
(\ref{Constants}) be generalized to other algebraic systems and
then used to study different integrable many body systems? The
answer of this questions goes beyond the scope of this paper and
will be considered elsewhere.

\begin{appendix}
\section{Obtaining the Eigenstates with Bethe Ansatz Method}

The simultaneous eigenstates and eigenvalues of the degenerate the
operators given in Eq. (\ref{Constants}) can be obtained using the
method of algebraic Bethe ansatz. In this method, one first
constructs a Bethe ansatz state \cite{Bethe:1931hc} which includes
some undetermined parameters and then substitutes this state into
the eigenvalue-eigenstate equation
$\hat{P}_j|\psi\rangle=E_j\psi\rangle$. The requirement that the
Bethe ansatz state obeys the eigenvalue-eigenstate equation yields
a set of equations called the equations of Bethe ansatz, whose
solutions determine the values of the parameters in the Bethe
ansatz state. For example, in order to obtain the eigenstates with
one pair of particles, one can start from a generic state in the
form
\begin{equation}\label{Eigenstate 1 x}
S^+(x)|0\rangle
\end{equation}
where $S^+(x)$ is defined in Eq. (\ref{Define S}). Using the
commutators given in Eq. (\ref{Quasispin Algebra}), one can show
that the action of the operator $\hat{P}_j$ on such a state is
given by
\begin{equation}\label{BA}
\hat{P}_jS^{+}(x)|0\rangle=
\left(E_j^{\left(0\right)}-\frac{\Omega_j}{1-c_j^{2}x}\right)S^{+}(x)|0\rangle
+
\left(x\sum_{j^\prime}\frac{\Omega_{j^\prime}}{1/c_{j^\prime}^2-x}\right)\frac{c_{j}S_j^{+}}{1-c_{j}^{2}x}|0\rangle.
\end{equation}
Clearly, if we choose $x$ in such a way that the second term on
the right hand side of Eq. (\ref{BA}) vanishes, i.e., if
\begin{equation}\label{BAE1}
x\sum_j\frac{\Omega_j}{1/c_j^2-x}=0,
\end{equation}
then Eq. (\ref{BA}) becomes an eigenvalue-eigenstate equation. One
way is to satisfy Eq. (\ref{BAE1}) is to take $x=0$ in which case
Eq. (\ref{BA}) yields the eigenvalue-eigenstate equation given in
Eq. (\ref{A1}). Alternatively one can choose $x$ in such a way
that it satisfies the Bethe ansatz equation
\begin{equation}
\sum_j\frac{\Omega_j}{1/c_j^2-x}=0
\end{equation}
in which case, Eq. (\ref{BA}) yields the eigenvalue-eigenstate
equation given in Eq. (\ref{A2}).

These results can be easily generalized to a state in the form
\begin{equation}\label{Eigenstate N x A}
S^+(x_1)S^+(x_2) \dots S^+(x_{N})|0\rangle
\end{equation}
which has $N$ pairs of particles where $N\leq N_{max}/2$. The
parameters $x_1,x_2,\dots,x_N$ are in general complex and they are
all different from one another. In fact, by acting on the state
given in Eq. (\ref{Eigenstate N x A}) with the operators
$\hat{P}_j$ one can easily show that if any two of the two
parameters $x_1,x_2,\dots,x_N$ are the same, then the state in Eq.
(\ref{Eigenstate N x A}) cannot be an eigenstate. If, on the other
hand, the parameters $x_1,x_2,\dots,x_N$ are all different from
one another, then by acting on the state given in Eq.
(\ref{Eigenstate N x A}) with the operators $\hat{P}_j$ given in
Eq. (\ref{Constants}), we find
\begin{eqnarray}\label{A6}
\hat{P}_{j}S^{+}\left( x_{1}\right) S^{+}\left( x_{2}\right)
\ldots S^{+}\left( x_{N}\right) |0\rangle 
&=&\left( E_{j}^{\left( 0\right) }-\sum_{k=1}^{N}\frac{\Omega
_{j}}{1-c_{j}^{2}x_{k}}\right) S^{+}\left( x_{1}\right)
S^{+}\left( x_{2}\right) \ldots S^{+}\left(
x_{N}\right) |0\rangle\\
&+&x_{1}\left( \sum_{\substack {k=1 \\ \left( k\not=1\right)}}
^{N}\frac{2}{x_{1}-x_{k}}+\sum_{{j^\prime}}\frac{\Omega
_{{j^\prime}}}{1/c_{{j^\prime}}^{2}-x_{1}}\right)\frac{c_{j}S_{j}^{+}}{1-c_{j}^{2}x_{1}}
S^{+}\left( x_{2}\right)
\ldots S^{+}\left( x_{N}\right) |0\rangle\nonumber\\
&+&x_{2}\left( \sum_{\substack {k=1 \\ \left( k\not=2\right)}}
^{N}\frac{2}{x_{2}-x_{k}}+\sum_{{j^\prime}}\frac{\Omega
_{{j^\prime}}}{1/c_{{j^\prime}}^{2}-x_{2}}\right) S^{+}\left(
x_{1}\right)
\frac{c_{j}S_{j}^{+}}{1-c_{j}^{2}x_{2}}\ldots S^{+}\left( x_{N}\right) |0\rangle \nonumber\\
&+&\ldots\nonumber\\
&+&x_{N}\left( \sum_{\substack {k=1 \\ \left( k\not=N\right)}}
^{N}\frac{2}{x_{N}-x_{k}}+\sum_{{j^\prime}}\frac{\Omega
_{{j^\prime}}}{1/c_{{j^\prime}}^{2}-x_{N}}\right) S^{+}\left(
x_{1}\right) S^{+}\left( x_{2}\right) \ldots
\frac{c_{j}S_{j}^{+}}{1-c_{j}^{2}x_{N}}|0\rangle\nonumber.
\end{eqnarray}
Clearly, Eq. (\ref{A6}) becomes an eigenvalue-eigenstate equation
for all the operators $\hat{P}_j$ if we choose the variables
$x_1$,$x_2$,$\dots$,$x_N$ so as to satisfy
\begin{equation}\label{BAE N}
x_k\left(\sum_{l=1(l\neq
k)}^N\frac{2}{x_k-x_l}+\sum_j\frac{\Omega_j}{1/c_j^2-x_k}\right)
=0
\end{equation}
for every $k=1,2,\dots,N$.

If the variables $x_1$,$x_2$,$\dots$,$x_N$ are all different from
zero, then they clearly have to obey
\begin{equation}
\sum_{l=1(l\neq
k)}^N\frac{2}{x_k-x_l}+\sum_j\frac{\Omega_j}{1/c_j^2-x_k} =0
\end{equation}
and in this case, Eq. (\ref{A6}) yields the eigenvalue-eigenstate
equation given in Eq. (\ref{A4}).

If, on the other hand, one of the parameters, say $x_1$, is chosen
to be zero (we cannot choose more than one $x_k$ to be zero since
the parameters must be different from one another) then the \Bae
~(\ref{BAE N}) is automatically satisfied for $k=1$. In this case,
the remaining parameters $x_2,\dots,x_N$ are to be found by
solving the $N-1$ equations
\begin{equation}\label{BAE N NZ}
\sum_j\frac{\Omega_j}{1/c_j^2-x_k}+\frac{2}{x_k}+\sum_{l=2(l\neq
k)}^N\frac{2}{x_k-x_l}=0
\end{equation}
for $k=2,3,\dots,N$. If we rename the remaining $N-1$ variables as
$x_2=z_1$, $x_3=z_2$, $\dots$, $x_N=z_{N-1}$, then Eq. (\ref{A6})
becomes the eigenvalue-eigenstate equation given in Eq.
(\ref{A3}).

\end{appendix}


\end{document}